\documentclass[aps,preprint,showpacs]{revtex4}
\usepackage{amssymb}
\usepackage{amsmath}
\usepackage{graphicx}
\usepackage{epsfig}
\usepackage{subfigure}
\usepackage{amsfonts}
\usepackage{txfonts}
\usepackage{CJK}
\usepackage{dcolumn}
\usepackage{bm}
\usepackage{cases}
\usepackage{color}
\usepackage[colorlinks,citecolor=blue, linkcolor=blue,hyperindex,CJKbookmarks,dvipdfm]{hyperref}
\begin{document}

\title{Waveguide transport mediated by strong coupling with atoms}

\author{Mu-Tian Cheng$^{1,2}$}
\email{mtcheng@ahut.edu.cn}
\author{Jingping Xu$^{1,3}$}
\author{Girish S. Agarwal$^{1}$}

\affiliation{$^{1}$Institute for Quantum Science and Engineering and Department of Biological and Agricultural Engineering,
Texas A\&M University, College Station, Texas 77845, USA}
\affiliation{$^{2}$School of Electrical Engineering $\&$ Information,
Anhui University of Technology, Maanshan 243002, P. R. China}
\affiliation{$^{3}$MOE Key Laboratory of Advanced Micro-Structured Materials, School of Physics Science and Engineering, Tongji University, Shanghai 200092, P. R. China}

\begin{abstract}
We investigate single photon scattering properties in one-dimensional waveguide coupled to quantum emitter's chain with dipole-dipole interaction (DDI). The photon transport is extremely sensitive to the location of the evanescently coupled atoms. The analytical expressions of reflection and transmission amplitudes for the chain containing two emitters with DDI are deduced by using real-space Hamiltonian. Two cases, where the two emitters symmetrically and asymmetrically couple to the waveguide, are discussed in detail. It shows that the reflection and transmission typical spectra split into two peaks due to the DDI. The Fano minimum in the spectra can be used to estimate the strength of the DDI. Furthermore, the DDI makes spectra strongly asymmetric and create a transmission window in the region where there was zero transmission. The scattering spectra for the chain consisting of multi-emitters are also given. Our key finding is that DDI can broaden the frequency band width for high reflection when the chain consists of many emitters.
\end{abstract}

\pacs{42.50.Nn, 42.50.Ct, 32.70.Jz}

\date{\today}

\maketitle

\section{\label{sec:1}Introduction}

Strong coupling between photons and atoms plays important roles in quantum information procession and quantum computation. Nanocavities, which can possess ultrasmall mode volume, are often used to realize the strong coupling \cite{Walther,reiserer}. Recently, both theoretical \cite{Chang1,hung} and experimental \cite{akimov,wei,huck,Babinec,Claudon,astafiev,hoi,yalla,avadi,Goban,sipahigil} works reported the strong coupling between the atoms and propagating photons in one-dimensional waveguide. Here, the strong coupling means that most of the energy from the atoms decays into the propagating modes of the waveguide. Based on the strong coupling, the photons scattering properties in one-dimensional waveguide are extensively investigated \cite{shen1,zhou1,liao,tsoi,tsoi2,shen2,shen3,witthaut,roy,longo,Zheng1,fang,cheng2,Bradford,neumeier,houla,xli,cheng3,burillo,Derouault,liaozeyang1,yanwb,yanch,Greenberg,kim1,kim} and reviewed in \cite{roy4}. Many quantum devices, such as single photon switching \cite{shen1,zhou1,liao,witthaut,cheng2,kim1,kim,liaozeyang1,yanwb,yanch}, router \cite{houla,xli,cheng3}, isolation \cite{xiakeyu}, transistor \cite{chang,neumeier,Kyriienko}, frequency comb generator \cite{zeyang},  and single photon frequency converter\cite{Bradford} have been proposed or realized. The one-dimensional waveguide can be photonic crystal waveguide \cite{Goban}, metal nanowire \cite{akimov,wei}, superconducting microwave transmission lines \cite{astafiev,hoi}, fiber \cite{yalla}, and diamond waveguide \cite{sipahigil}. Atoms and cavities can play the role of scatter. Atomic chain is also an important scatter. The coupling between one-dimensional waveguide and atomic chain can lead to many interesting phenomena, such as superradiant decays \cite{Goban}, and changing optical band structure \cite{albrecht}. It can also be used to realize Bragg mirrors \cite{corzo,sorensen} and single photon isolator \cite{isola}.\par

It is well-known that if the separation between two atoms is much smaller than the resonance wavelength, the dipole-dipole interaction (DDI) can be strong \cite{agarwal}. It has been shown that the DDI can change the single photon scattering properties\cite{tianwei,yu,cheng}. But in these studies, the two atoms are localized in one cavity \cite{tianwei,cheng} or the same place along the waveguide \cite{yu}. The spatial separation between the two atoms along the waveguide direction are not involved. However, the separation plays important role in some important phenomena, such as quantum beats \cite{zhenghaix}, generation entanglement \cite{chengy,Gonzalez,Facchi,mirza}, single photon switching \cite{kim} and Bragg mirrors \cite{corzo,sorensen}. Recently, Liao \textsl{et al.} investigated the time evolution of emitter excitations and photon pulse in the one-dimensional waveguide coupled to multiple emitters with DDI \cite{liaoddi}. In this paper, we study the single photon scattering properties by using the real-space Hamiltonian. The analytical expressions for reflection and transmission amplitudes for the case of two quantum emitters (QEs) with DDI are given. The single photon scattering properties for many QEs with DDI are also exhibited. The results show that the DDI can significantly affect the single photon scattering properties. \par

The structure of this paper is organized as follows. In Sec. \uppercase\expandafter{\romannumeral2}, we present the model of single photon transport in a waveguide coupled to atoms. In Sec. \uppercase\expandafter{\romannumeral3}, we recall the known results for the case of a single atom. In Sec.  \uppercase\expandafter{\romannumeral4}, we present new features of the two atom coupling. In Sec. \uppercase\expandafter{\romannumeral5}, we discuss the trend for many atoms with conclusions in Sec. \uppercase\expandafter{\romannumeral6}.

\section{\label{sec:2}Model and Hamiltonian}

\begin{figure}[ptb]
\includegraphics[width=7cm
]{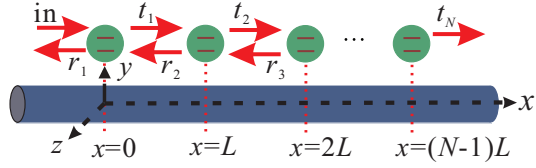} \caption{(Color online) The system considered in the manuscript. A chain of $N$ QEs with equal separation $L$ coupled to one-dimensional waveguide.} \label{fig1}
\end{figure}

The system considered in this paper is shown in Fig.\ref{fig1}. $N$ QEs with equal separation $L$ side couple to a waveguide. The QEs are modeled as two-level systems with ground state $|g\rangle$ and excited state $|e\rangle$.  The transition frequency of the QEs is $\omega_{A}$.

When $\omega_{A}$ is much larger than the cutoff frequency $\omega_{c}$ of the waveguide, the dispersion relation of the waveguide near the resonant frequency can be taken as linear \cite{tsoi}. Then the Hamiltonian in the real space is given by $H=H_{f}+H_{i}+H_{d}$, with $\hbar=1$, \cite{tsoi,chengy}
\begin{equation}
H_{f}=iv_{g}\int dx(a^{\dag}_{L}(x)\frac{\partial a_{L}(x)}{\partial x}-a^{\dag}_{R}(x)\frac{\partial a_{R}(x)}{\partial x})
+\sum^{N}_{j=1}(\omega_{A}-i\Gamma^{'}_{0j}/2)\sigma^{(j)}_{ee},
\end{equation}
being the free propagation photon in the waveguide and the QEs. $v_{g}$ is the group velocity of the photon. $a^{\dag}_{R}(x) (a^{\dag}_{L}(x))$ means creation a right (left) propagation photon at $x$. $\sigma^{(j)}_{ee}=|e\rangle_{j}\langle e|$. We have supposed that all the transition frequencies of the atoms are the same and the energy of QE's ground state is zero.  $\Gamma^{'}_{0j}$ is the energy decay rate into the non-waveguide modes.
\begin{equation}
H_{i}=\sum^{N}_{j=1}J_{j}\int dx \{\delta(x-x_{j}){[a^{\dag}_{R}(x)+a^{\dag}_{L}(x)] \sigma_{j}+H.c.\}},
\end{equation} denotes the interaction between the QEs and the waveguide photon. $J_{j}$ is the couple strength between the $j$-th QE and the waveguide photon. $\sigma_{j}=|g\rangle_{j}\langle e|$ is the ladder operator for the $j$-th QE. Finally,
\begin{equation}
H_{d}=\Omega_{i,j}\sum^{N}_{i,j=1}(\sigma^{\dag}_{i}\sigma_{j}+\sigma^{\dag}_{j}\sigma_{i})
\end{equation}
describes the DDI. $\Omega_{ij}=\frac{3}{4}\Gamma_{0}[(\frac{\cos x}{x^{3}}+\frac{\sin x}{x^{2}}-\frac{\cos x}{x})+\cos^{2}\theta (\frac{\cos x}{x}-\frac{3\cos x}{x^{3}}-\frac{3\sin x}{x^{2}})]$ is the DDI strength between the $i$-th QE and the $j$-th QE \cite{agarwal}.  $x\equiv\frac{\omega_{A}}{c}|\vec{r}_{i}-\vec{r}_{j}|$. $\cos^{2}\theta=(\frac{\vec{p}\cdot(\vec{r}_{i}-\vec{r}_{j})}{|p|\cdot|\vec{r}_{i}-\vec{r}_{j}|})^{2}$, where $\vec{r}_{j}$ is the location coordinate of the $j$-th QE and $\vec{p}$ is the dipole. We suppose that all the QEs have the same dipoles and their directions are all in $-y$. $\Gamma_{0}$ is the decay rate of QE in free space, which is taken about 7.5 MHz in the following calculations.\par

Since only one exciation exists in this system, the eigenstate of $H$ takes the form
\begin{equation}
|E_{k}\rangle=\int dx[\phi_{kR}(x)a^{\dag}_{R}(x)+\phi_{kL}(x)a^{\dag}_{L}(x)]|0,g\rangle+\sum^{N}_{j=1}e^{(j)}_{k}|0,e_{j}\rangle ,
\end{equation}
where $E_{k}=\hbar \omega_{k}$ is the eigenvalue of $H$. $|0,g\rangle $ represents all the QEs in the ground state and no photon in the system. $|0,e_{j}\rangle $ denotes no photon in the system and the $j$-th QE in the excited state $|e\rangle $ while all other QEs in the ground state. $e^{(j)}_{k} $ is the probability amplitude of the state $|0,e_{j}\rangle $. $\phi_{kR}(x) $ and $\phi_{kL}(x) $ are the amplitudes of the fields going to the right and left in the waveguide. These are continuous except at the positions of the atoms and thus we write these in the form \cite{tsoi,tsoi2}
\begin{numcases}{\phi_{kR}(x)=}
e^{ikx} \quad (x<0),\nonumber \\
t_{j}e^{ik(x-jL)}   \quad  (j-1)L<x<jL\label{eq5}\\
t_{N}e^{ik(x-NL)}     \quad x>(N-1)L,\nonumber
\end{numcases}
and

\begin{numcases}{\phi_{kL}(x)=}
r_{1}e^{-ikx} \quad (x<0),\nonumber \\
r_{j+1}e^{-ik(x-jL)}   \quad  (j-1)L<x<jL,\label{eq6}\\
0,    \quad x>(N-1)L,\nonumber
\end{numcases}
where $t_{j}$ and $r_{j}$ are the coefficients to be determined. Substituting Eqs.(\ref{eq5}) and (\ref{eq6}) into the Schr\"{o}dinger equation $H|E_{k}\rangle=E_{k}|E_{k}\rangle$, we obtain \cite{tsoi}

\begin{subequations}
\begin{align}
&t_{j}e^{-ikL}-t_{j-1}+\frac{iJ_{j}e^{(j)}_{k}}{v_{g}}=0,\label{7a}\\
&r_{j+1}e^{ikL}-r_{j}-\frac{iJ_{j}e^{(j)}_{k}}{v_{g}}=0, \\
&t_{j-1}+r_{j}+\frac{\sum^{j-1}_{i=1}\Omega_{j,i}e^{(i)}_{k}+\sum^{N}_{i=j+1}\Omega_{j,i}e^{(i)}_{k}}{J_{j}}-\frac{(\Delta^{(j)}_{k}+i\Gamma^{'}_{0j}/2) e^{(j)}_{k}}{J_{j}}=0,\label{7c}
\end{align}
\end{subequations}
where $\Delta^{(j)}_{k} =\omega_{k}-\omega^{(j)}_{A}$. $t_{0}=1$ and $r_{N+1}=0$ are used in the following calculations. Clearly from Eqs.(\ref{eq5}) and (\ref{eq6}), the transmission and reflection coefficients will be $t=t_{N}e^{-ikNL}$ and $r=r_{1}$, respectively.

\section{single emitters}
Before showing how DDI affects on the single photon scattering properties, we review single photon scattered by one QE first. The single photon transmission and reflection amplitudes are, respectively, given by \cite{shen1,xli}
\begin{subequations}
\begin{align}
t=\frac{\Delta_{k}+i\Gamma^{'}_{0}/2}{i\Gamma+\Delta_{k}+i\Gamma^{'}_{0}/2},\label{sigat}\\
r=\frac{-i\Gamma}{i\Gamma+\Delta_{k}+i\Gamma^{'}_{0}/2},\label{sigar}
\end{align}
\end{subequations}
where, $\Delta_{k} =\omega_{k}-\omega_{A}$ and $\Gamma=J^{2}/v_{g}$. Eqs. (\ref{sigat}) and (\ref{sigar}) show that reflection probability $R\equiv|r|^{2}$ reaches the maximum and transmission probability $T\equiv|t|^{2}$ reaches the minimum when $\Delta_{k}=0$. \par

The blue solid lines in Fig. \ref{fig2} exhibit the numerical results $T$ and $R$ with different coupling strengths between the QE and the photon in the nanowaveguide. In the numerical model, a semiconductor quantum dot (QD) with resonant wavelength $\lambda_{qd}=655$ nm (transition frequency $\omega_{A}/(2\pi)\approx$ 457.7 THz) is placed near a Ag nanowire, which was realized in experiments \cite{akimov}. The radius of Ag nanowire is 10 nm. And the corresponding wavelength of the propagating surface plasmom (SP) $\lambda_{sp}$ is about 211.8 nm \cite{Chang1}, which is much shorter than the resonant wavelength of the QD due to the reduced group velocity. The spontaneous emission rate $\Gamma_{pl}$ into the propagation surface plasmon modes, the energy losses rate $\Gamma^{'}_{0}$, which consists of radiating into the free space rate $\Gamma_{rad}$ and non-radiative emission rate into the Ag nanowire $\Gamma_{non-rad}$, is calculated by using the formulas given in Ref.\cite{Chang1}. \par

\section{two quantum emitters}
\subsection{Symmetric coupling}
We now show how the DDI affects on the single photons scattering properties for the case of a pair of QEs coupling to the waveguide. First, we discuss the results for the symmetric coupling ($J_{1}=J_{2}=J$). From Eqs. (\ref{7a}) to (\ref{7c}), one can obtain the analytical solutions to the $t$ and $r$, which are given by
\begin{subequations}
\begin{align}
t=\frac{e^{-ikL}\{-i\Gamma\Omega+ie^{2ikL}\Gamma\Omega+e^{ikL}[(\Delta_{k}+i\Gamma^{'}_{0}/2)^{2}-\Omega^{2}]\}}{(-1+e^{2ikL})\Gamma^{2}+2i\Gamma(\Delta_{ k}+i\Gamma^{'}_{0}/2+e^{ikL}\Omega)+(\Delta_{k}+i\Gamma^{'}_{0}/2)^{2}-\Omega^{2}},\label{9a}\\
r=\frac{(1-e^{2ikL})\Gamma^{2}-i\Gamma[(1+e^{2ikL})(\Delta_{k}+i\Gamma^{'}_{0}/2)+2e^{ikL}\Omega]}{(-1+e^{2ikL})\Gamma^{2}+2i\Gamma(\Delta_{ k}+i\Gamma^{'}_{0}/2+e^{ikL}\Omega)+(\Delta_{k}+i\Gamma^{'}_{0}/2)^{2}-\Omega^{2}}\label{9b},
\end{align}
\end{subequations}
where $\Omega$ is the DDI strength between the two QEs. When $\Omega=0$, which means that DDI is not considered, one can obtain

\begin{subequations}
\begin{align}
t=\frac{(\Delta_{k}+i\Gamma^{'}_{0}/2)^{2}}{(-1+e^{2ikL})\Gamma^{2}+2i\Gamma(\Delta_{ k}+i\Gamma^{'}_{0}/2)+(\Delta_{k}+i\Gamma^{'}_{0}/2)^{2}},\label{symit}\\
r=\frac{(1-e^{2ikL})\Gamma^{2}-i\Gamma(1+e^{2ikL})(\Delta_{k}+i\Gamma^{'}_{0}/2)}{(-1+e^{2ikL})\Gamma^{2}+2i\Gamma(\Delta_{ k}+i\Gamma^{'}_{0}/2)+(\Delta_{k}+i\Gamma^{'}_{0}/2)^{2}}\label{symir},
\end{align}
\end{subequations}
which is consistent with previous reports \cite{zhenghaix,cheng4}.

Note that we can write the denominator in Eq. (\ref{symit}) as $(\Delta_{k}+\frac{i\Gamma^{'}_{0}}{2}+i\Gamma)^{2}+\Gamma^{2}e^{2ikL}$. The term $\Gamma^{2}e^{2ikL}$ arises from the waveguide mediated interactions between two QEs even if the direct DDI  $\Omega=0$. Thus it plays the role of waveguide mediated DDI. To show this, if we drop $\Gamma^{2}e^{2ikL}$, then transmission $t^{(2)}$ for two QE case is the square of the transmission $t^{(1)}$ for the single QE case. Thus in the absence of $\Gamma^{2}e^{2ikL}$ term, the field transmitted by the first QE is transmitted by the second QE leading to the result $t^{(2)}=(t^{(1)})^{2}$. However, the physics is different. The field reflected by the second QE affects the first QE changing its transmission which then changes the fields produced by the second QE. In principle, one has whole series of such processes and this just happens to be the physics of DDI. Hence we refer to the term $\Gamma^{2}e^{2ikL}$ as waveguide mediated DDI. We have checked that this DDI affects line shapes but does not produce splitting.\par
\begin{figure}[ptb]
\includegraphics[width=14cm
]{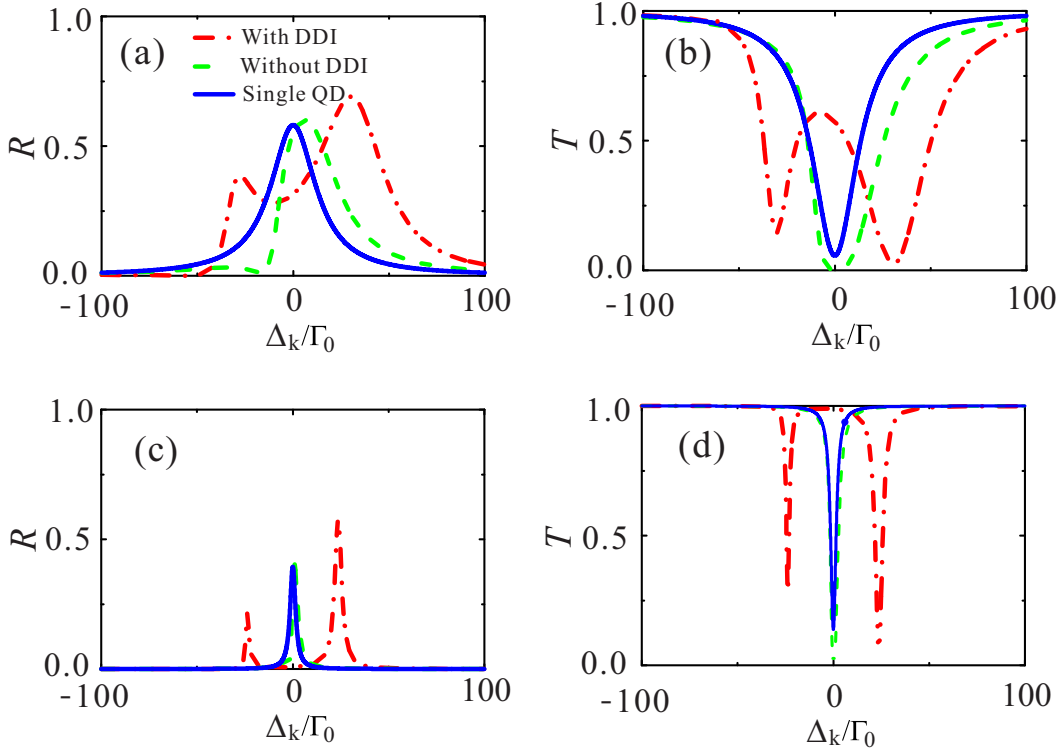} \caption{(Color online) $R$ and $T$ as a function of $\Delta_{k}$. The blue solid lines are the results for single QDs. The red dashed-dotted lines denote the results for a pair of QDs with DDI and the green dashed lines for a pair of QDs without DDI. In (a) and (b), the two QDs locates at $\vec{r}_{1}(x, y, z)$=(0, 17 nm, 0) and  $\vec{r}_{2}(x,y,z)$=(32.75 nm, 17 nm, 0), respectively, corresponding to the separation between the two QDs $L=\lambda_{qd}/20$. $\Gamma$=11.03$\Gamma_{0}$ and $\Gamma^{'}_{0}$= 6.86$\Gamma_{0}$  are used in the calculations. In (c) and (d),  the two QDs are placed at $\vec{r}_{1}(x, y, z)$=(0, 37 nm, 0) and  $\vec{r}_{2}(x, y, z)$=(32.75 nm, 37 nm, 0 ), respectively. $\Gamma=1.06\Gamma_{0}$ and $\Gamma^{'}_{0}=1.26\Gamma_{0}$. The single QD is located at (0,17 nm,0) and (0,37 nm, 0) when the blue solid lines are plotted in (a,b) and (c,d), respectively. In the calculations, $\Omega=23.08\Gamma_{0}$.
}\label{fig2}
\end{figure}

\begin{figure}[ptb]
\includegraphics[width=8cm
]{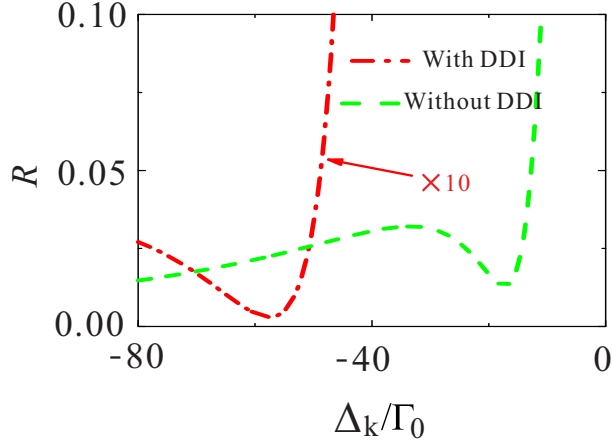} \caption{(Color online) Fano-shape of the reflection spectrum, [the region of minimum in Fig.\ref{fig2}(a)].}\label{fig3coo}
\end{figure}

\begin{figure}[ptb]
\includegraphics[width=8cm
]{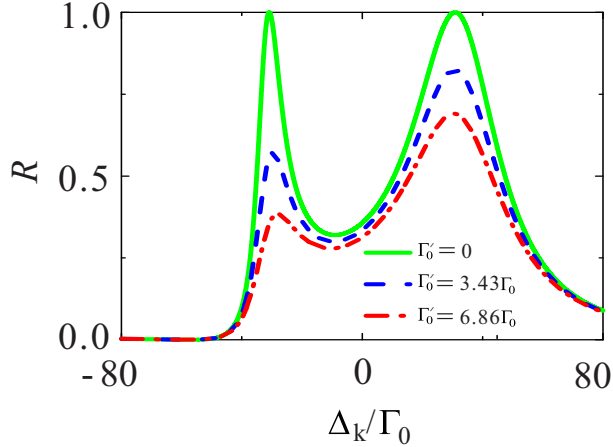} \caption{(Color online) Reflection spectra for the two identical QDs with different decays. In the calculations, $\Gamma$=11.03$\Gamma_{0}$, $L=32.75$ nm and $\Omega=23.08\Gamma_{0}$.}\label{fig44}
\end{figure}

The transmission and reflection spectra for the case of $|\vec{r}_{1}-\vec{r}_{2}|=\lambda_{qd}/20$ are shown in Fig.\ref{fig2}. Here, $kL\approx0.31\pi$, which is due to the short wavelength of SP. Without considering the DDI, the
reflection spectrum reaches the maximum at $\Delta_{k}=0$ and Fano-lineshape appears \cite{cheng4}. Compared to the single QE case, the two QE spectra display considerable asymmetries even in the absence of DDI.  The position of the Fano minimum can be estimated from the zero of the  numerator in Eq. (\ref{9b}). For $\Gamma^{'}_{0}=0$, it is found to occur at $\Delta_{k}=\Delta^{rmin}_{k}=-\Gamma(\tan kL+\frac{\Omega}{\Gamma}\sec kL)$, which depends on the strength of DDI. Clearly, the position of the Fano minimum can be used to get an estimate of the DDI strength  and this is displayed more clearly in Fig.\ref{fig3coo}. The distance between the two dips in Fig.\ref{fig3coo} is related to $\Omega\sec kL$. The Fano-lineshapes in the reflection spectra are reduced strongly but still exist.
Eqs. (\ref{9a}) and (\ref{9b}) also give that the single photon reflection spectrum splits into two main peaks at $\Delta_{k}=\Delta^{rmax}_{k}=\pm \sqrt{2\Gamma\Omega\sin(kL)+\Omega^{2}}$ when $\Gamma^{'}_{0}=0$. These values give the positions where $R=1,T=0$. Note that $\Delta^{rmax}_{k}$ depends on both DDI and $kL$ but DDI is absolutely essential for $\Delta^{rmax}_{k}\neq0$. These also correspond to the positions of dips in the transmission spectrum. The differences of the main peaks of the reflection spectrum in Fig.\ref{fig2} result from the energy losses. To show this claim clearly, we present Fig.\ref{fig44}, which shows that when energy loss is zero, both of the two peaks reach the maximum of one. However, when energy loss increases from 3.43$\Gamma_{0}$ to 6.86$\Gamma_{0}$, the difference between the heights of the two peaks increases from about 0.25 to about 0.31. In the numerical calculations, we take $kL=(\omega_{A}+\Delta_{k})/v_{g}\approx2\pi L/\lambda_{sp}$ since $\omega_{A}\gg\Gamma_{0}$ and $\Delta_{k}$ \cite{kim,tsoi} . \par

\begin{figure}[ptb]
\includegraphics[width=8cm
]{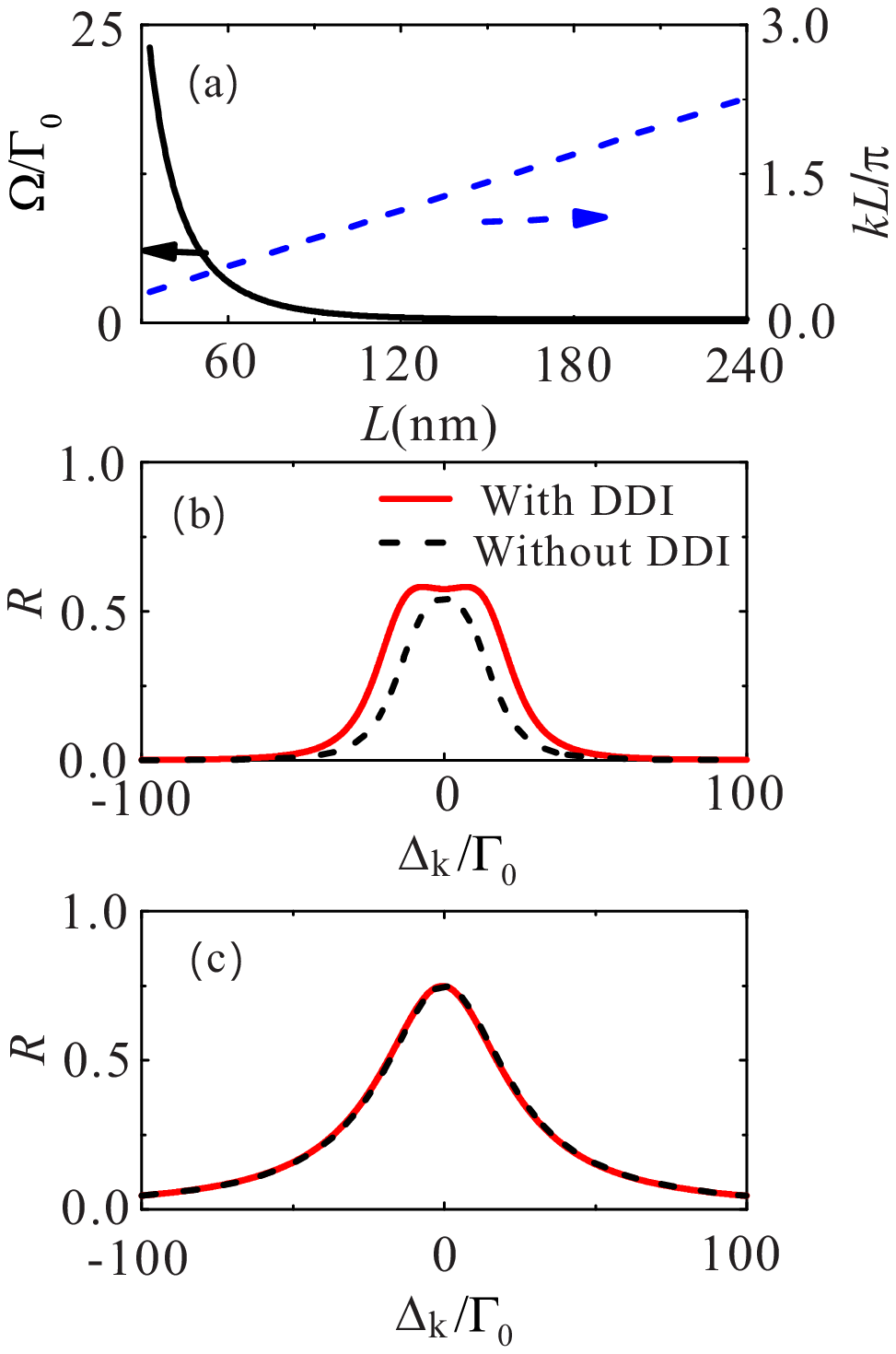} \caption{(Color online) (a) DDI strength as a function of the distance $L$ between the two QDs, i.e. dots located $(x,y,z)$ and $(x+L,y,z)$. (b) and (c) are the single photon reflection spectra for the case of $L=\lambda_{sp}/4=52.95$ nm and $L=\lambda_{sp}/2=105.9$ nm. In (b), $\vec{r}_{1}(x,y,z)$=(0,17 nm,0), $\vec{r}_{2}(x,y,z)$=(52.95 nm,17 nm,0), $\Omega=5.12\Gamma_{0}$. In (c) $\vec{r}_{1}(x,y,z)$=(0, 17 nm,0), $\vec{r}_{2}(x,y,z)$=(105.9 nm, 17 nm, 0), $\Omega=0.61\Gamma_{0}$. In both (b) and (c), $\Gamma$=11.03$\Gamma_{0}$ and $\Gamma^{'}_{0}$= 6.86$\Gamma_{0}$. }\label{fig4}
\end{figure}
Many reports show that single photon scattering properties and their applications such as single photon switching, generation entanglement, are strongly related on the distance between two QEs.  The DDI strength also depends strongly on the distance between the two QEs. Fig. \ref{fig4}(a) shows $\Omega$ as a function of $L$ for the two QDs with resonant wavelength $\lambda_{qd}=655$ nm. The $\Omega$ decreases from $23.08\Gamma_{0}$ to $0.28\Gamma_{0}$ as $L$ increasing from 32.75 nm to 240 nm. Fig. \ref{fig4}(b) and (c) exhibit single photon reflection spectra for $L=52.95$ nm (corresponding to $L=\lambda_{sp}/4, kL=\pi/2$) and $L=105.9$ nm (corresponding to $L=\lambda_{sp}/2, kL=\pi$), respectively. It indicates that DDI can play significant roles even though the separation between the two QDs reaches $L=\lambda_{sp}/4$. However, when the distance increases to $L=\lambda_{sp}/2$, the influence of DDI can be neglected, \textsl{i.e.}, the numerical results with and without DDI are almost indistinguishable but not identical. \par

It is need to point out that spatial separation between the two QDs along the waveguide direction plays important role in splitting the reflection spectrum. There are special cases when the DDI yields a shift in spectrum rather than splitting. This happens when the numerator and denominator share a common zero. As an example if $kL=0$, which can be realized, for example, the two QDs located at $\vec{r}_{1}(x,y,z)=(0,0,37$ nm), $\vec{r}_{2}(x,y,z)=(0,37$ nm, 0), respectively, then $t=(\Delta_{k}-\Omega+i\Gamma^{'}_{0}/2)/((\Delta_{k}-\Omega+i\Gamma^{'}_{0}/2)+2i\Gamma)$, and $r=-2i\Gamma/((\Delta_{k}-\Omega+i\Gamma^{'}_{0}/2)+2i\Gamma)$. The collective behavior is still present as the effective line width parameter is changed from $\Gamma$ to 2$\Gamma$. There is no splitting in the reflection spectrum but the location of the peak in the reflection spectrum shifts to $\omega_{k}=\omega_{A}+\Omega$. A similar result is obtained for $kL=\pi$, one needs to replace $\Omega$ by $-\Omega$. Thus the relative phase factor $kL$ produced by the propagation of the light from QE 1 and to QE 2 is important in the transport of light through a waveguide coupled to QEs.\par

\subsection{Asymmetric coupling}
\begin{figure}[ptb]
\includegraphics[width=14cm
]{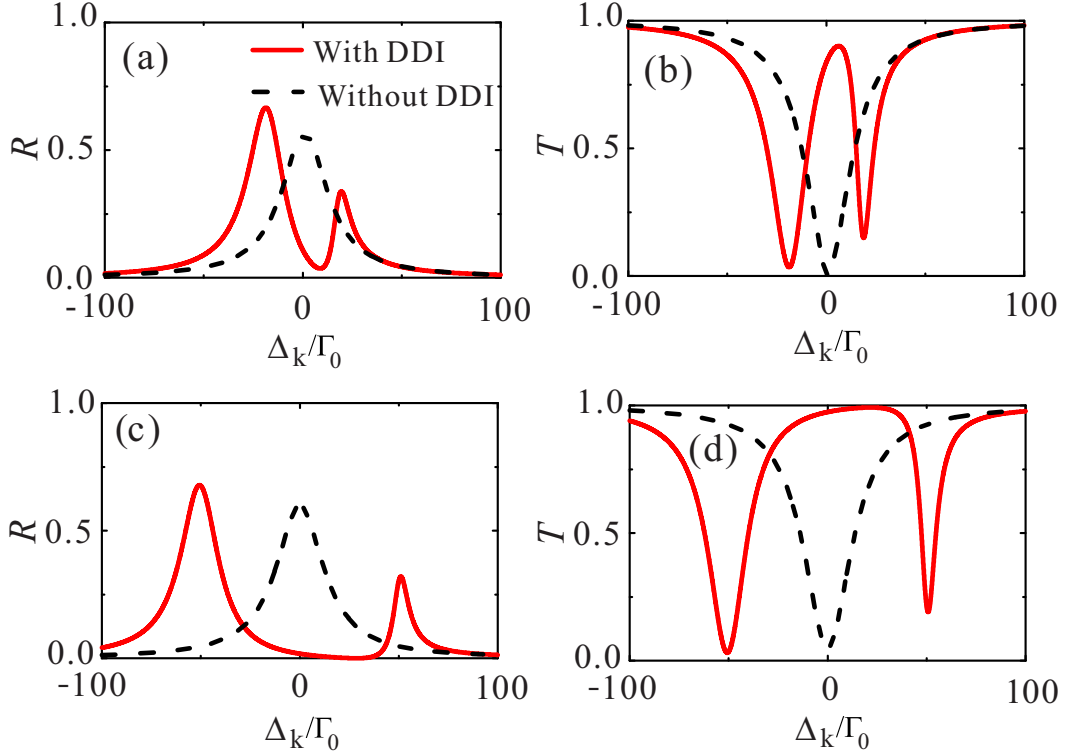} \caption{(Color online) The single photon reflection and transmission spectra for the two QDs placed in different locations. In (a) and (b), $\vec{r}_{1}(x,y,z)$=(0, 17 nm, 0), $\vec{r}_{2}(x,y,z)$=(20 nm, 37 nm, 0), corresponding to $kL=0.19\pi$. $\Gamma_{1}$=11.03$\Gamma_{0}$ and $\Gamma^{'}_{01}$= 6.86$\Gamma_{0}$,  $\Gamma_{2}=1.06\Gamma_{0}$, $\Gamma^{'}_{02}=1.26\Gamma_{0}$ and $\Omega=-20.79\Gamma_{0}$. In (c) and (d), $\vec{r}_{1}(x,y,z)$=(0, 17 nm, 0), $\vec{r}_{2}(x,y,z)$=(0, 49.75 nm, 0), corresponding to $kL=0$.  $\Gamma_{1}$=11.03$\Gamma_{0}$ and $\Gamma^{'}_{01}$= 6.86$\Gamma_{0}$,  $\Gamma_{2}=0.33\Gamma_{0}$, $\Gamma^{'}_{02}=1.12\Gamma_{0}$ and $\Omega=-50.71\Gamma_{0}$.}\label{fig5}
\end{figure}

We now discuss the single photon scattering with asymmetric coupling. If the distances between the surface of the nanowire and the two QDs are different, then the coupling strengths $J_{1}\neq J_{2}$. From Eqs. (\ref{7a}) to (\ref{7c}), one can get
\begin{subequations}
\begin{align}
t=\frac{e^{-ikL}\{-i\sqrt{\Gamma_{1}\Gamma_{2}}\Omega+ie^{2ikL}\sqrt{\Gamma_{1}\Gamma_{2}}\Omega+e^{ikL}(\delta^{(1)}_{k}\delta^{(2)}_{k}-\Omega^{2})\}}
{(-1+e^{2ikL})\Gamma_{1}\Gamma_{2}+i(\Gamma_{1}\delta^{(2)}_{k}+\Gamma_{2}\delta^{(1)}_{k})+2ie^{ikL}\sqrt{\Gamma_{1}\Gamma_{2}}\Omega+\delta^{(1)}_{k}\delta^{(2)}_{k}-\Omega^{2}},\label{10a}\\
r=\frac{(1-e^{2ikL})\Gamma_{1}\Gamma_{2}-ie^{2ikL}\Gamma_{2}\delta^{(1)}_{k}-i\Gamma_{1}\delta^{(2)}_{k}-2ie^{ikL}\sqrt{\Gamma_{1}\Gamma_{2}}\Omega}
{(-1+e^{2ikL})\Gamma_{1}\Gamma_{2}+i(\Gamma_{1}\delta^{(2)}_{k}+\Gamma_{2}\delta^{(1)}_{k})+2ie^{ikL}\sqrt{\Gamma_{1}\Gamma_{2}}\Omega+\delta^{(1)}_{k}\delta^{(2)}_{k}-\Omega^{2}}\label{10b},
\end{align}
\end{subequations}
where $\Gamma_{j}=J^{2}_{j}/v_{g}, \delta_{j} =\Delta_{k}+i\Gamma^{'}_{0j}/2 (j=1,2)$. Eqs.(\ref{10a}) and (\ref{10b}) indicate that when $\Delta^{2}_k=\sqrt{\Gamma_{1}\Gamma_{2}}\Omega\sin(kL)+\Omega^{2}$ is satisfied, $T=0, R=1$ if $\Gamma^{'}_{0j}=0$. Furthermore, if $L=0$, which can be realized for the two QDs with the same location coordinates $x, z$ but different $y$, the condition changes to be $\Delta^{2}_k=\Omega^{2}$. This means that the distance between the two peaks in the reflection spectrum is not only dependent on the coupling strength via $\Gamma$ but also on $\Omega$. However, if $\Omega=0$, Eqs. (\ref{10a}) and (\ref{10b}) degenerate into
\begin{subequations}
\begin{align}
t=\frac{\delta^{(1)}_{k}\delta^{(2)}_{k}}
{(-1+e^{2ikL})\Gamma_{1}\Gamma_{2}+i(\Gamma_{1}\delta^{(2)}_{k}+\Gamma_{2}\delta^{(1)}_{k})+\delta^{(1)}_{k}\delta^{(2)}_{k}},\label{12a}\\
r=\frac{(1-e^{2ikL})\Gamma_{1}\Gamma_{2}-ie^{2ikL}\Gamma_{2}\delta^{(1)}_{k}-i\Gamma_{1}\delta^{(2)}_{k}}
{[(-1+e^{2ikL})\Gamma_{1}\Gamma_{2}+i(\Gamma_{1}\delta^{(2)}_{k}+\Gamma_{2}\delta^{(1)}_{k})+\delta^{(1)}_{k}\delta^{(2)}_{k}}\label{12b}.
\end{align}
\end{subequations}
There is no splitting in the reflection spectrum. Fig.(\ref{fig5}) shows $R$ and $T$ as a function of $\Delta_{k}$ for asymmetric coupling. Both the cases of $kL\neq0$ and $kL=0$ are shown. It exhibits clearly that the DDI splits the reflection and transmission spectrum.\par
\section{multi Quantum Emitters}
\begin{figure}[ptb]
\includegraphics[width=14cm
]{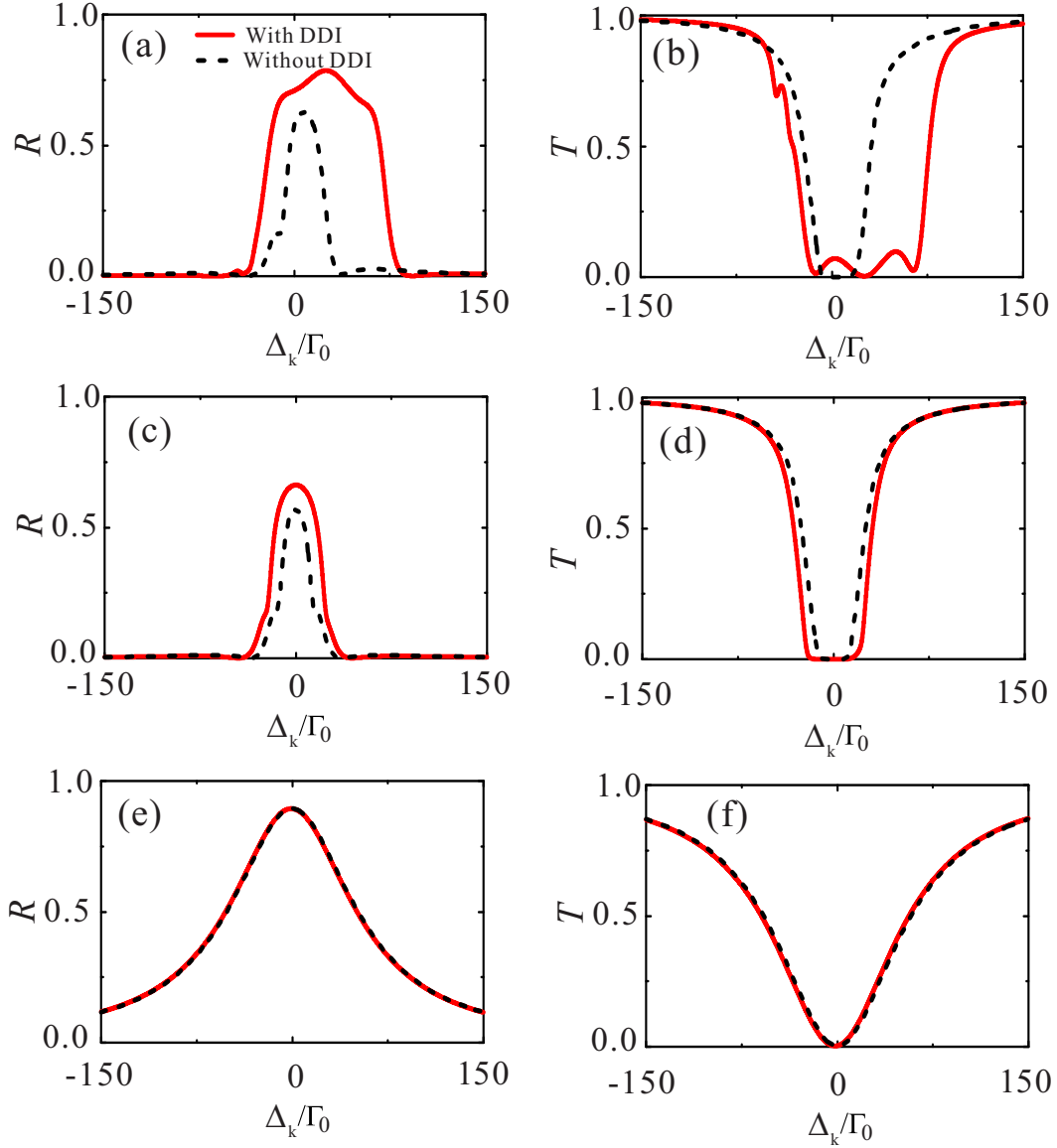} \caption{(Color online) The single photon reflection and transmission spectra for the case of 5 QDs coupled to the nanowire. The separations between two neighbouring QDs along $x$ direction are $\lambda_{qd}/20$=32.75 nm (a,b),  $\lambda_{sp}/4$=52.95 nm (c,d), and $\lambda_{sp}/2$=105.9 nm (e,f). In the calculations, $\Gamma$=11.03$\Gamma_{0}$ and $\Gamma^{'}_{0}$= 6.86$\Gamma_{0}$. The DDI coupling for each pair has been calculated using Fig \ref{fig4}(a). As an example, in (a), the coupling between first and second is 23.08$\Gamma_{0}$ and between first and third is 2.60$\Gamma_{0}$.}\label{fig6}
\end{figure}

If one QE couples to the waveguide, only the photon with the frequency equaling to the transition frequency of the QE reflects perfectly. Based on the coupled-resonator waveguide, Chang \textsl{et al}. proposed using many atoms individually in the resonators to realize perfect reflection of single photon in a wide band of frequency \cite{changmulti}. Here, we exhibit that the DDI can broaden the band width. Fig. \ref{fig6} shows the single photon reflection spectra where 5 QDs couple to the Ag nanowire. The separations between the two neighbouring QDs are 32.75 nm (a,b),  52.95 nm (c,d) and 105.9 nm (e,f), where the distance dependent $\Omega$ is considered, which can be found in Fig.\ref{fig4}(a). In Fig. \ref{fig6}(a), the width is broaden about 2.5 times than that without DDI. Furthermore, the maximum of the peak in the reflection spectra is also enhanced. Fig. \ref{fig5}(c) shows that the band width of the reflection spectrum is broadened and the peak is increased even though $L$ reaches $\lambda_{sp}/4 (kL=\pi/2)$. However, if $L$ further increases to $\lambda_{sp}/2 (kL=\pi)$, the numerical results with and without DDI are almost indistinguishable, as shown in Fig. \ref{fig6}(e).\par
\begin{figure}[ptb]
\includegraphics[width=14cm
]{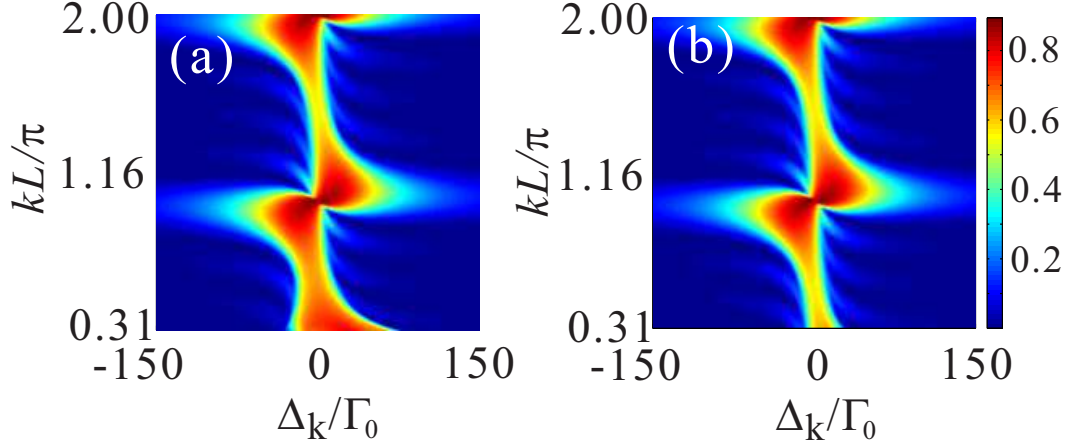} \caption{(Color online) The single photon reflection spectra as a function of $\Delta_{k}$ and $kL$ with (a) and without (b) DDI. In the calculations, $\Gamma$=11.03$\Gamma_{0}$ and $\Gamma^{'}_{0}$= 6.86$\Gamma_{0}$. The DDI coupling has been calculated using Fig \ref{fig4}(a).}\label{fig8}
\end{figure}

To exhibit how DDI broadens reflection spectra clearly, we present $R$ as a function of $\Delta_{k}$ and $kL$ with and without DDI in Fig. \ref{fig8}. When $kL\approx0.31\pi$, $\Omega$ is about $23.08\Gamma_{0}$. It can affect the scattering spectrum strongly, as we discussed above. It shows that the red region in the direction of $\Delta_{k}$ in Fig. \ref{fig8}(a) is much larger than in Fig. \ref{fig8}(b). However, when $kL$ increases to more than $\pi$, the influence of DDI can be neglected, then there is no obvious difference between Fig. \ref{fig8}(a) and (b).
\section{Conclusions}
In summary, we have investigated single photon scattering properties in one-dimensional waveguide coupled to an array QEs with DDI by using real-space Hamiltonian. For the case of the chain consisting of two QEs with symmetric coupling, the reflection spectrum splits into two peaks due to the DDI, however, the splitting depends on both the DDI coupling and spatial separation between the two QEs along the photon propagation direction in waveguide. The spectra also display the Fano interference minimum. With two QEs, there are new pathways which lead to transmission and reflection. For example a new pathway will consist of-the radiation after being scattered by QE1 interacts with QE2; the scattered radiation from QE2 interacts back with QE1. Thus the transmitted wave has additional contribution from this path way. The new pathways result in Fano minimum. For both symmetric and asymmetric couplings, the DDI can induce reflection spectrum splitting. The distance between the two peaks in the reflection spectrum depends on the DDI strength strongly in both symmetric and asymmetrical couplings case. DDI can also broaden the frequency band width of high reflection probability of single photon when many QEs couple to the waveguide. Our results may find applications in design single photon devices and quantum information processing.

\begin{acknowledgments}
MTC acknowledges discussions with Dr. Zeyang Liao, and support from the Anhui Provincial Natural Science Foundation under Grant Nos.1608085MA09, 1408085QA22 and China Scholarship Council. GSA thanks the Biophotonics initiative of the Texas A$\&$M University for support.
\end{acknowledgments}

\end{document}